\documentclass[prd,amsmath,amssymb,superscriptaddress,nofootinbib]{revtex4}

\usepackage{amsfonts}
\usepackage{multirow}
\usepackage{makecell}
\usepackage{mathrsfs}
\usepackage{graphicx}
\usepackage{amsmath}
\usepackage{amssymb}
\usepackage{bm}
\usepackage{bbm}
\usepackage{color}
\usepackage{ulem}
\usepackage{array}
\usepackage{diagbox}
\usepackage{float}
\allowdisplaybreaks[4] 

\def\OMIT#1{}

\def\hlinew#1{%
  \noalign{\ifnum0=`}\fi\hrule \@height #1 \futurelet
   \reserved@a\@xhline}
\usepackage{array}
\newcommand{\PreserveBackslash}[1]{\let\temp=\\#1\let\\=\temp}
\newcolumntype{C}[1]{>{\PreserveBackslash\centering}p{#1}}
\newcolumntype{R}[1]{>{\PreserveBackslash\raggedleft}p{#1}}
\newcolumntype{L}[1]{>{\PreserveBackslash\raggedright}p{#1}}

\newcommand{\beq}{\begin{equation}}
\newcommand{\eeq}{\end{equation}}
\newcommand{\bqa}{\begin{eqnarray}}
\newcommand{\eqa}{\end{eqnarray}}

\newcommand{\jpsi}{J/\psi}

\newcommand\fverb{\setbox\fverbbox=\hbox\bgroup\verb}
\newcommand\fverbdo{\egroup\medskip\noindent%
            \fbox{\unhbox\fverbbox}\ }
\newcommand\fverbit{\egroup\item[\fbox{\unhbox\fverbbox}]}
\newbox\fverbbox

\makeatletter

\newcommand{\Rmnum}[1]{\expandafter\@slowromancap\romannumeral #1@}
\allowdisplaybreaks[2]

\makeatother

\begin{document}


\title{Optimized ${\cal O}(\alpha_s^2)$ correction to exclusive
double $J/\psi$ production at $\bm{B}$ factories}


\author{Wen-Long Sang~\footnote{wlsang@swu.edu.cn}}
\affiliation{School of Physical Science and Technology, Southwest University, Chongqing 400700, China\vspace{0.2cm}}

\author{Feng Feng\footnote{f.feng@outlook.com}}
\affiliation{Institute of High Energy Physics and Theoretical Physics Center for Science Facilities, Chinese Academy of Sciences, Beijing 100049, China\vspace{0.2cm}}
\affiliation{China University of Mining and Technology, Beijing 100083, China\vspace{0.2cm}}

\author{Yu Jia\footnote{jiay@ihep.ac.cn}}
\affiliation{Institute of High Energy Physics and Theoretical Physics Center for Science Facilities, Chinese Academy of Sciences, Beijing 100049, China\vspace{0.2cm}}
\affiliation{School of Physics, University of Chinese Academy of Sciences, Beijing 100049, China\vspace{0.2cm}}

\author{Zhewen Mo\footnote{mozw@ihep.ac.cn }}
\affiliation{Institute of High Energy Physics and Theoretical Physics Center for Science Facilities, Chinese Academy of Sciences, Beijing 100049, China\vspace{0.2cm}}
\affiliation{School of Physics, University of Chinese Academy of Sciences, Beijing 100049, China\vspace{0.2cm}}

\author{Jichen Pan\footnote{panjichen@ihep.ac.cn }}
\affiliation{Institute of High Energy Physics and Theoretical Physics Center for Science Facilities, Chinese Academy of Sciences, Beijing 100049, China\vspace{0.2cm}}
\affiliation{School of Physics, University of Chinese Academy of Sciences, Beijing 100049, China\vspace{0.2cm}}

\author{Jia-Yue Zhang\footnote{zhangjiayue@ihep.ac.cn}}
\affiliation{Institute of High Energy Physics and Theoretical Physics Center for Science Facilities, Chinese Academy of Sciences, Beijing 100049, China\vspace{0.2cm}}
\affiliation{School of Physics, University of Chinese Academy of Sciences, Beijing 100049, China\vspace{0.2cm}}

\date{\today}

\begin{abstract}
The failure of observing the $e^+ e^- \to J/\psi+J/\psi$ events at $B$ factories to date
is often attributed to the significant negative order-$\alpha_s$ correction.
In this work we compute the ${\cal O}(\alpha^2_s)$ correction to this process for the first time.
The magnitude of the next-to-next-to-leading order (NNLO) perturbative correction is substantially negative
so that the standard NRQCD prediction would suffer from an unphysical, negative cross section.
This dilemma may be traced in the fact that the bulk contribution of the fixed-order radiative corrections stems from the
perturbative corrections to the $J/\psi$ decay constant.
We thus implement an improved NRQCD factorization framework, by decomposing the amplitude into the photon-fragmentation piece and
the non-fragmentation piece. With the measured $J/\psi$ decay constant as input, which amounts to resumming a specific class of radiative and
relativistic corrections to all orders, the fragmentation-induced production rate can be predicted accurately and serves a benchmark prediction.
The non-fragmentation type of the amplitude is then computed through NNLO in $\alpha_s$ and at lowest order in velocity.
Both the ${\cal O}(\alpha_s)$ and ${\cal O}(\alpha^2_s)$ corrections in the interference term become positive and exhibit a decent convergence behavior.
Our finest prediction is $\sigma(e^+ e^- \to J/\psi+J/\psi)=  2.13^{+0.30}_{-0.06}$ fb  at $\sqrt{s}=10.58$ GeV.  With
the projected integrated luminosity of 50 ${\rm ab}^{-1}$, the prospect to observe this exclusive
process at \texttt{Belle} 2 experiment appears to be bright.
\end{abstract}

\maketitle
\section{introduction}

The exclusive production of hadrons in $e^+e^-$ collider with center-of-mass energy far below $Z$ pole is dominated
by $e^+e^-$ annihilation into a single virtual photon, in which the final-state hadrons must have an overall
negative $C$ parity. These types of processes are best exemplified by the exclusive double charmonium production processes $e^+e^-\to\jpsi+\eta_c$ and
$\jpsi+\chi_{cJ}$. The experimental observation of these channels at $B$ factories in the beginning of this century~\cite{Belle:2002tfa,BaBar:2005nic}
has stimulated tremendous theoretical efforts, where the majorities of investigations were conducted 
within the non-relativistic QCD (NRQCD) factorization framework~\cite{Braaten:2002fi, Liu:2002wq, Hagiwara:2003cw, Zhang:2005cha, Gong:2007db, He:2007te, Bodwin:2007ga, Dong:2012xx, Xi-Huai:2014iaa,  Feng:2019zmt, Huang:2022dfw,Zhang:2008gp, Wang:2011qg, Dong:2011fb, Jiang:2018wmv, Sun:2021tma, Sang:2022kub}.

In 2006 \texttt{BarBar} collaboration reported two exclusive processes about production of
two neutral vector mesons, and the measured production rates are
$\sigma (e^+ e^-\to \rho^0 \rho^0)=20.7 \pm 0.7(\text{stat})\pm 2.7(\text{syst})$ fb and $\sigma (e^+ e^-\to   \rho^0 \phi )=5.7 \pm 0.5(\text{stat})\pm0.8(\text{syst})$ fb,
with a cut $|\cos{\theta}|<0.8$ imposed~\cite{BaBar:2006vxk}. Since the final state has net positive $C$ parity, these exclusive processes must proceed via $e^+e^-$ annihilation into
two photons. The suppression caused by extra QED coupling constants can be largely compensated by the significant enhancement
brought by the small photon virtuality once the vector mesons are produced
through two photon independent fragmentation, hence the production rates can be surprisingly larger than naively expected.
Shortly after, Davier, Peskin and Snyder~\cite{Davier:2006fu} (see also Bodwin {\it et al.}~\cite{Bodwin:2006yd}) considered these processes
in the vector dominance model (VMD). Including the finite width of $\rho^0$, these authors obtained
$\sigma (e^+ e^-\to \rho^0 \rho^0)=21.4 \pm 0.7$ fb and $\sigma (e^+ e^-\to \rho^0 \phi )=6.15 \pm 0.22$ fb
with $|\cos{\theta}|<0.8$, in satisfactory agreement with the \texttt{BarBar} measurements~\cite{BaBar:2006vxk}.

One naturally speculates whether the similar yet much cleaner double $J/\psi$ production process can be observed at $B$ factories or not.
In fact, as early as in 2003, \texttt{Belle} experiment has already looked for this channel and not found clear signal~\cite{Belle:2004abn}.
Instead an upper limit is placed, $\sigma(e^+ e^- \to  J/ \psi J/ \psi )\mathcal{B}_{>2}< 9.1$ fb at the 90$\%$ confidence level,
where $\mathcal{B}_{>2}$ signifies the branching fraction for final states
including more than two charged tracks.

On the theoretical side, the $e^+ e^- \to  J/ \psi J/ \psi$ process has already been investigated by several different groups over the years.
In 2002 Bodwin, Braaten, Lee studied this process at lowest order in NRQCD approach and predicted the cross section to be around
$8.7$ fb~\cite{Bodwin:2002fk},  which is even greater than the LO NRQCD prediction for $e^+e^-\to J/\psi+\eta_c$.
Shortly after, this prediction has been updated to $6.65$ fb by the same authors~\cite{Bodwin:2002kk}.
With the aid of VMD, Davier, Peskin, Snyder considered the photon fragmentation contribution only
and predicted the total cross section about $2.38$ fb~\cite{Davier:2006fu}.
Besides the photon fragmentation contribution, Bodwin, Braaten, Lee and Yu further took into account the non-fragmentation contribution
within the NRQCD factorization framework, and found a sizable destructive interference effect, with the
cross section predicted to be about $1.69\pm 0.35$ fb~\cite{Bodwin:2006yd}.

An important progress was made by Gong and Wang in 2008~\cite{Gong:2008ce}, who computed the ${\cal O}(\alpha_s)$ correction to this process,
yet at the lowest order in velocity. The NLO perturbative correction turns out to be negative and significant.
By including the radiative correction, Gong and Wang found that the LO prediction about $7.4\sim 9.1$ fb reduces to $-3.4\sim 2.3$ fb~\cite{Gong:2008ce}.
Later the combined NLO perturbative and relativistic corrections were investigated by Fan, Lee and Yu in 2012~\cite{Fan:2012dy}.
They found the fixed-order NRQCD prediction for the cross section to range from $-12$ fb to $-0.43$ fb,
which is negative and sensitive to the charm quark mass and renormalization scale.  On the other hand,
following the recipe practised in \cite{Bodwin:2006yd}, splitting the amplitude into the photon-fragmentation and
non-fragmentation parts,  Fan, Lee and Yu found the predicted cross section boosted to the positive range
$1 \sim 1.5$ fb~\cite{Fan:2012dy}.

The predicted double-$J/\psi$ cross sections at $B$ factories are scattered in  a wide range.
To provide useful guidance for experimentalists to search for this channel,
it is crucial to make the most precise theoretical prediction, which is the chief motivation of this work.
Since the ${\cal O}(\alpha_s)$ correction is quite important, one naturally wonders what is the impact of ${\cal O}(\alpha^2_s)$ correction.
It is the goal of this work to investigate the two-loop QCD correction to this double-$J/\psi$ production process,
which turns out to be an exceedingly formidable task.
It is found that the standard NRQCD approach leads to a substantially negative ${\cal O}(\alpha^2_s)$ perturbative correction,
so that the predicted cross section would become negative, thus unphysical.
We may attribute this symptom to the fact that, the bulk contribution of the fixed-order perturbative
corrections to this process largely arises from the perturbative corrections to the $J/\psi$ decay constant,
which are also negative and sizable.
Motivated by this observation, we follow \cite{Bodwin:2006yd} to adopt an optimized NRQCD factorization approach,
by decomposing the production amplitude into the photon-fragmentation and non-fragmentation pieces.
The fragmentation-initiated production rate can be predicted unambiguously with the measured $J/\psi$ decay constant as input.
The interference and the non-fragmentation parts are then computed through NNLO in $\alpha_s$ within NRQCD approach.
In this improved NRQCD framework, we find that both the ${\cal O}(\alpha_s)$ and ${\cal O}(\alpha^2_s)$ corrections in the interference part
become positive and exhibit a good convergence behavior.
Our finest prediction in the optimized NRQCD approach turns out to be quite close to the fragmentation contribution.
We believe that our NNLO prediction from the improved NRQCD approach
is much more robust and reliable than that from the traditional NRQCD approach, and we hope that the
\texttt{Belle} 2 experiment will examine our prediction in the near future.

The rest of the paper is distributed as follows.
In Sec.~\ref{Fragmentation:production:rate}, we recapitulate the photon fragmentation contribution to $e^+e^-\to J/\psi J/\psi$.
In Sec.~\ref{Strategy:Improved:NRQCD}, we introduce the improved NRQCD factorization approach,  versus the traditional NRQCD factorization.
In Sec.~\ref{LO:Improved:NRQCD}, we present the tree-level prediction for the cross section in the improved NRQCD scheme.
In Sec.~\ref{Higher:order:prediction:improved:NRQCD}, we outline the procedure of computing
the NLO and NNLO perturbative corrections to this process in the improved NRQCD approach,
present the numerical results and explore phenomenological consequence.
In Sec.~\ref{two:loop:NRQCD:examination}, we take a class of two-gluon-exchange two-loop diagrams as example,
to illustrate the structure of
the infrared poles and examine their cancellation.
The purpose is to exhibit that the validity of NRQCD factorization is highly nontrivial.
Finally we summarize in Sec.~\ref{Summary}.
For the sake of completeness, in Appendix~\ref{appendix:A}, we present the NLO and NNLO predictions for the
$e^+e^-\to J/\psi J/\psi$ process in the standard NRQCD factorization approach.

\section{Exclusive production of double $J/\psi$ via two photon independent fragmentation}
\label{Fragmentation:production:rate}

\begin{figure}[h!]
\centering
\includegraphics[scale=0.7]{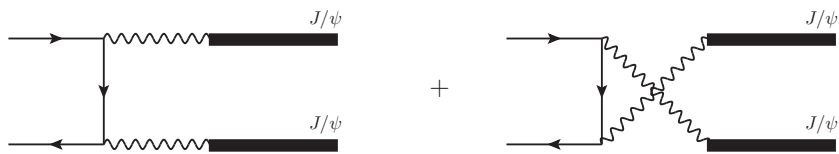}
\caption{Illustration of the $e^+e^-\to J/\psi+J/\psi$ process through two photon independent fragmentation. }
\label{diagrams:2Jpsi:from:photon:fragmentation}
\end{figure}

Since the double-$J/\psi$ in the final state has the even $C$-parity, this production process must proceed through $e^+e^-$ annihilation into
two virtual photons. The dominant production mechanism is via two photon independent fragmentation into $J/\psi$, as shown in
Fig.~\ref{diagrams:2Jpsi:from:photon:fragmentation}. According to the vector dominance model (VMD),
the photon-to-$J/\psi$ coupling strength is governed by $e e_c M_{J/\psi} f_{J/\psi}$,
where $f_{J/\psi}$ denotes the $J/\psi$ decay constant and is defined by $\langle J/\psi|\bar{c}\gamma^\mu c|0\rangle=-f_{J/\psi} M_{J/\psi}
\varepsilon_{J/\psi}^{*\mu}$. The value of the decay constant can be determined from the
precisely measured leptonic width of $J/\psi$:
\beq
\Gamma(J/\psi\to l^+l^-)= {4\pi e_c^2 \alpha^2 \over 3} {f_{J/\psi}^2\over M_{J/\psi}},
\label{lepton:width:Jpsi:fV}
\eeq
where $\alpha$ denotes fine-structure constant and $e_c =\frac{2}{3}$  is the electric charge of the charm quark.

The differential unpolarized cross section for $e^+e^-\to J/\psi J/\psi$ through photon fragmentation has been computed by Davier, Peskin and Synder in 2006~\cite{Davier:2006fu}
(see also \cite{Bodwin:2006yd}):
\beq
{d\sigma_{\rm fr}(e^+e^-\to J/\psi J/\psi) \over d\cos{\theta}} =
\left(\frac{e e_c f_{J/\psi}}{M_{J/\psi}}\right)^4 {\pi \alpha^2\over s} \beta \frac{(t^2+u^2)(tu-M_{J/\psi}^4)+4stuM_{J/\psi}^2}{t^2 u^2 },
\label{diff:X:section:photon:fragmentation}
\eeq
with $\beta = \sqrt{1- 4M_{J/\psi}^2/s}$ represents the velocity of the outgoing $J/\psi$.
in the center-of-mass frame. This expression is symmetric under $t\leftrightarrow u$ due to Bose symmetry.
In the high energy limit $\sqrt{s}\gg M_{J/\psi}$,  \eqref{diff:X:section:photon:fragmentation} reduces to
\beq
{d\sigma_{\rm fr}(e^+e^-\to J/\psi J/\psi) \over d\cos{\theta}} \approx  \left( {e e_c f_{J/\psi}\over M_{J/\psi}} \right)^4
{d\sigma(e^+e^-\to \gamma\gamma ) \over d\cos{\theta}}.
\eeq

Integrating \eqref{diff:X:section:photon:fragmentation} over $\cos\theta$ (one should cover only the hemisphere of the solid angle since two $J/\psi$
are indistinguishable bosons), one obtains
\beq
\sigma_\mathrm{fr} = \frac{32 \pi ^3 e_c^4 \alpha ^4 f_{J/\psi}^4 }{M_{J/\psi}^4 } {1\over s} \left[\frac{4+(1-\beta^2)^2}{1+\beta^2}
\ln \left(\frac{1+\beta}{1-\beta}\right)-2\beta \right].
\label{integrated:X:section:photon:fragmentation}
\eeq
Note the fragmentation-initiated total cross section exhibits the asymptotic $1/s$ decrease, which is identical to the scaling behavior of
$e^+e^-\to {\rm hadrons}$ and in sharp contrast with the $1/s^4$ scaling associated with the exclusive
process $e^+e^-\to J/\psi+\eta_c(\chi_{c1})$ and the $1/s^3$ scaling affiliated with  $e^+e^-\to J/\psi+\chi_{c0,2}$.

Plugging the latest \texttt{BESIII} measurement of $\Gamma(J/\psi\to l^+l^-)=5.56$ keV~\cite{BESIII:2022wsp},
$M_{J/\psi} = 3.0969$ GeV, and the running QED coupling $\alpha\left(M_{J/\psi}\right)=1/132.642$ into \eqref{lepton:width:Jpsi:fV},
one obtains $f_{J/\psi} = 403$ MeV.
Substituting this value and $\sqrt{s}=10.58$ GeV in \eqref{integrated:X:section:photon:fragmentation},
one then obtains $\sigma_\mathrm{fr} = 2.52$ fb.

\section{Strategy of improved NRQCD prediction for $e^+e^-\to J/\psi J/\psi$}
\label{Strategy:Improved:NRQCD}

\subsection{Traditional NRQCD factorization prediction}

The mainstream theoretical tool to account for the exclusive charmonium production nowadays is the
NRQCD factorization approach~\cite{Bodwin:1994jh}. This approach allows one to express the cross section as a double expansion in $\alpha_s$
and $v$, the typical charm quark velocity inside $J/\psi$. Concretely speaking, at the lowest order in $v$,
the NRQCD prediction for the production rate of
$e^+e^-\to J/\psi J/\psi$ can be cast in the following form:
\beq
\frac{d\sigma}{d\cos\theta}=\frac{1}{2s}\frac{\beta}{16\pi} \frac{e^8e_c^4}{4}\bigg( {\cal F}^{(0)}
+\frac{\alpha_s}{\pi} {\cal F}^{(1)}+ \left(\frac{\alpha_s}{\pi}\right)^2{\cal F}^{(2)} +\cdots \bigg)
{\vert \langle {\cal O}\rangle_{J/\psi}\vert^2 \over m_c^2}+{\cal O}(v^2),
\label{Conventional:NRQCD:factorization:formula}
\eeq
where ${\cal F}^{(i)}$ ($i=0,1,2$) represent the short-distance coefficients (SDCs) at various perturbative order,
and $\langle {\cal O}\rangle_{J/\psi}$ is the abbreviation of the following NRQCD matrix element:
\beq
\langle {\mathcal O} \rangle_{J/\psi} \equiv
\vert \langle J/\psi(\lambda)|\psi^{\dagger}\boldsymbol{\sigma}\cdot\boldsymbol{\varepsilon}(\lambda)\chi|0\rangle\vert^2.
\eeq
In Appendix~\ref{appendix:A} we will present the numerical predictions through NNLO in $\alpha_s$ within the traditional NRQCD factorizatoin framework.

\subsection{Improved NRQCD factorization prediction}

At any prescribed order in $\alpha_s$, the double $J/\psi$ production from $e^+e^-$ annihilation proceeds through either the photon fragmentation or
non-fragmentation channel, where the former always dominates the latter.
Following \cite{Bodwin:2006yd}, we split the production amplitude into the fragmentation and non-fragmentation pieces:
\beq
\frac{d\sigma}{d\cos\theta} = \frac{1}{2s}\frac{\beta}{16\pi} \frac{1}{4}\sum_{\rm spin} \left\vert {\cal M}_{\rm fr}+
{\cal M}_{\rm nfr} \right\vert^2.
\label{dsigma:dcos:decomposition}
\eeq
One then applies NRQCD factorization to the non-fragmentation part of the amplitude, which is expressed in terms of the charm quark mass and
$\langle {\cal O}\rangle_{J/\psi}$, rather than $M_{J/\psi}$ and $f_{J/\psi}$.
After squaring the amplitude in \eqref{dsigma:dcos:decomposition} and summing over spins,
we decompose the differential cross section into the
fragmentation part, interference part and the non-fragmentation part:
\beq
\frac{d\sigma}{d\cos\theta} = \frac{1}{2s}\frac{\beta}{16\pi} \frac{e^8 e_c^4}{4}\bigg[\mathcal{C}_{\rm fr}f_{J/\psi}^4+\mathcal{C}_{\rm int} f_{J/\psi}^2 \frac{
\langle {\cal O} \rangle_{J/\psi}}{m_c}+\mathcal{C}_{\rm nfr}\bigg(\frac{\langle {\cal O} \rangle_{J/\psi}}{m_c}\bigg)^2\bigg].
\label{Optimized:NRQCD:factorization:formula}
\eeq

The coefficient affiliated with the fragmentation piece can be read off from
\eqref{diff:X:section:photon:fragmentation}:
\beq
  \mathcal{C}_{\text{fr}}=\frac{8\left(\left(t^2+u^2\right)(tu-M_{J/\psi}^4)+4stuM_{J/\psi}^2\right)}{t^2u^2M_{J/\psi}^4}.
\label{diff:X:sec:frag}
\eeq

The interference and nonfragmentation terms can be tackled in NRQCD factorization approach.
At lowest order in $v$ but through ${\alpha_s^2}$, the coefficients can be parameterized as
\begin{subequations}
\bqa
\mathcal{C}_{\rm int} &=& {\cal C}_{\rm int}^{(0)}\bigg[1+\frac{\alpha_s}{\pi}\hat{c}_{\rm int}^{(1)}+\left(\frac{\alpha_s}{\pi}\right)^2
\bigg(\frac{\beta_0}{4}\ln\frac{\mu_R^2}{m_c^2}\hat{c}_{\rm int}^{(1)}+2\gamma_{J/\psi}\ln\frac{\mu_\Lambda^2}{m_c^2}+\hat{c}_{\rm int}^{(2)}\bigg)+\cdots\bigg],
\label{C:int:parametrization}
\\
\mathcal{C}_{\rm nfr}&=& {\cal C}_{\rm nfr}^{(0)}\bigg[1+\frac{\alpha_s}{\pi}\hat{c}_{\rm nfr}^{(1)}+\left(\frac{\alpha_s}{\pi}\right)^2
\bigg(\frac{\beta_0}{4}\ln\frac{\mu_R^2}{m_c^2}\hat{c}_{\rm nfr}^{(1)}+4\gamma_{J/\psi}\ln\frac{\mu_\Lambda^2}{m_c^2}+\hat{c}_{\rm nfr}^{(2)}\bigg)+\cdots \bigg],
\label{C:nfr:parametrization}
\eqa
\label{C:int:nfr:param}
\end{subequations}
where $\mu_R$ and $\mu_\Lambda$ refer to renormalization scale and NRQCD factorization scale. $\beta_0=11C_A/3-2 n_f/3$, with $n_f=4$ signifying the number
of active quark flavors. The occurrence of the $\beta_0\ln \mu_R$ term is dictated by the renormalization group invariance.
$\gamma_{J/\psi}=-\frac{\pi^2}{12}C_{F}(2C_F+3C_A)$ is the two-loop anomalous dimension of the NRQCD vector current~\cite{Czarnecki:1997vz,Beneke:1997jm}.
The occurrence of the $\gamma_{J/\psi}\ln \mu_\Lambda$ term at two-loop order is demanded by the NRQCD factorization.
$\hat{c}^{(i)}$ ($i=1,2$) represent the non-logarithmic order-$\alpha_s$ and order-$\alpha_s^2$ corrections.

\begin{figure}[h]
    \centering
    \includegraphics[scale=0.8]{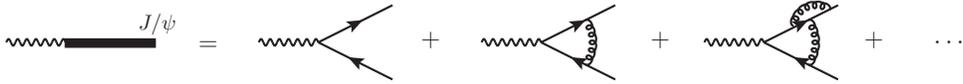}
    \caption{Diagrammatic illustration of the NRQCD factorization for the $J/\psi$ decay constant.}
    \label{Illustration:Decay:Constant:NRQCD:fac}
\end{figure}

Both $f_{J/\psi}$ and $\langle {\mathcal O} \rangle_{J/\psi}$ enter the improved NRQCD factorization
formula \eqref{Optimized:NRQCD:factorization:formula}. However, these two nonperturbative parameters are actually interrelated.
Unlike the decay constants of light vector mesons, $f_{J/\psi}$ is not an entirely nonperturbative object and rather encapsulates some
perturbative effect.  As picturised in Fig.\ref{Illustration:Decay:Constant:NRQCD:fac},
NRQCD factorization allows one to further factorize the $J/\psi$ decay constant as short-distance coefficients multiplied with the
NRQCD matrix element $\langle {\mathcal O} \rangle_{J/\psi}$:
\beq
f_{J/\psi} = \sqrt{2 \langle {\mathcal O} \rangle_{J/\psi} \over M_{J/\psi}}\left[1+ \mathfrak{f}^{(1)}\frac{\alpha_s}{\pi}+\left(\frac{\alpha_s}{\pi}\right)^2
\left(\mathfrak{f}^{(1)}\frac{\beta_0}{4}\ln \frac{\mu_R^2}{m_c^2}+\gamma_{J/\psi}\ln\frac{\mu_{\Lambda}^2}{m_c^2}+ \mathfrak{f}^{(2)}\right) +\cdots \right]
+{\cal O}(v^2),
\label{decay:constant:NRQCD:fac}
\eeq
with $\mathfrak{f}^{(1)}=- 2 C_F$, $\mathfrak{f}^{(2)}=-43.3288$~\cite{Czarnecki:1997vz,Beneke:1997jm}, and $n_f =4$.
The ${\cal O}(\alpha_s^3)$ correction~\cite{Marquard:2014pea,Feng:2022vvk} and ${\cal O}(\alpha^i_s v^2)$ ($i=0,1$) corrections~\cite{Keung:1982jb,Luke:1997ys}
have also been available.

We adopt the measured value $f_{J/\psi}=403$ MeV in \eqref{Optimized:NRQCD:factorization:formula}.
This implies that we have resummed an infinite towers of perturbative and relativistic corrections to all orders.
To avoid double counting, we must exclude those fragmentation-type diagrams in our analysis
at any given perturbative order.

\section{Leading-order cross section in improved NRQCD factorization}
\label{LO:Improved:NRQCD}

\begin{figure}[h!]
    \centering
    \includegraphics[scale=1.0]{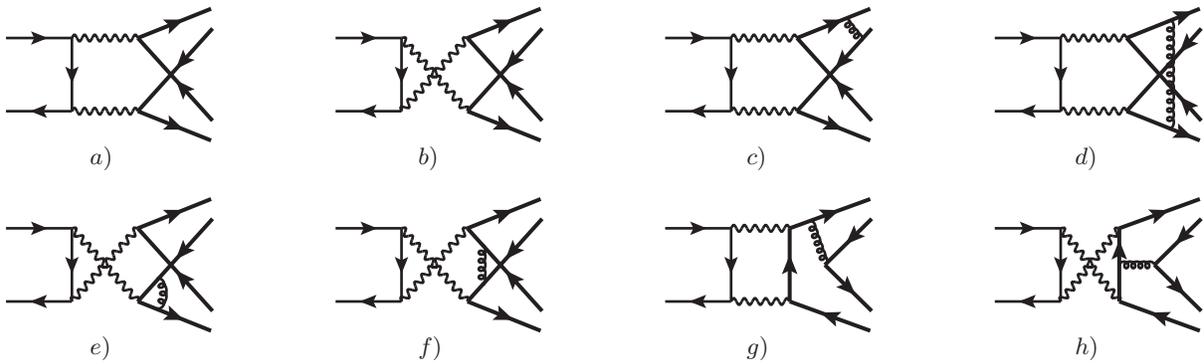}
    \caption{Non-fragmentation type of tree-level Feynman diagrams [$a)$ and $b)$], together with
    some sample one-loop non-fragmentation diagrams [$c)$ through $h)$]. }
    \label{Fig:nonfrag:diagrams:LO:NLO}
\end{figure}

There are two tree-level non-fragmentation diagrams, as depicted in Fig.~\ref{Fig:nonfrag:diagrams:LO:NLO}$a)$ and $b)$.
A straightforward calculation yields the tree-level coefficients for the interference and the non-fragmentation terms in \eqref{C:int:nfr:param}:
\begin{subequations}
\bqa
{\cal C}_{\text{int}}^{(0)} &=& -\frac{128 \left(t u(t^2+u^2)+20m_c^2 s t u +16m_c^4 s^2-64m_c^6 s-512m_c^8\right)}{3 m_c^2\, t\, u\, s^3},
\\
{\cal C}_{\text{nfr}}^{(0)}&=& 2048\left(\frac{-12t u(t u-32m_c^4+4m_c^2s)+5s^2 t u}{9 s^6}
+\frac{16m_c^2(s^3-5m_c^2s^2+48m_c^4s-192m_c^6)}{9 s^6}\right).
\eqa
\label{c0:int:nfr}
\end{subequations}

Substituting \eqref{diff:X:sec:frag} and  \eqref{c0:int:nfr} into \eqref{Optimized:NRQCD:factorization:formula},
integrating over $\cos\theta$ from 0 to 1, we reproduce the fragmentation-initiated integrated cross section
\eqref{integrated:X:section:photon:fragmentation}, and obtain
the following interference and non-fragmentation contributions to the integrated cross section:
\begin{subequations}
\begin{align}
& \sigma_\mathrm{int} = -\frac{16 \pi ^3  e_c^4 \alpha ^4 f_{J/\psi}^2 \langle {\cal O} \rangle_{J/\psi}}{3 m_c^3 s^2} \left[
(5-\beta^2)(1-\beta^2)^2\ln \left(\frac{1+\beta}{1-\beta}\right)
+22 \beta -{40\over 3}\beta^3 + 2 \beta^5\right],
\\
& \sigma_\mathrm{nfr} = \frac{2048 \pi ^3 \alpha ^4 e_c^4 | \langle {\cal O} \rangle _{J/\psi}|^2}{45m_c^2 s^3}
\beta \left( 10- {20\over 3}\beta^2+\beta^4\right).
\end{align}
\label{total:X:section:tree:int:nfr}
\end{subequations}
Here the $J/\psi$ velocity $\beta$ is evaluated by replacing $M_{J/\psi}$ with $2 m_c$.

In contrast with the fragmentation part that asymptotically scales as $1/s$,
the interference part of the cross section exhibits a $1/s^2$ asymptotic decrease,  while the non-fragmentation part
exhibits a $1/s^3$ scaling.
Adding  \eqref{integrated:X:section:photon:fragmentation} and \eqref{total:X:section:tree:int:nfr},
setting $f_{J/\psi} \approx \sqrt{\langle {\cal O}\rangle_{J/\psi}/m_c}$ and $M_{J/\psi}\approx 2 m_c$ everywhere,
we reproduce the analytic expression of the tree-level integrated cross section~\cite{Gong:2008ce}.

In phenomenological analysis, one often approximates the NRQCD matrix element by
\beq
\langle {\cal O}\rangle_{J/\psi} \approx \frac{3}{2\pi} R^2_{J/\psi}(0),
\eeq
where $R_{J/\psi}(0)$ represents the radial Schr\"{o}dinger wave function of $J/\psi$ at the origin.

In this work, we choose to use $R^2_{J/\psi}(0)=0.81\:{\rm GeV^3}$ from Buchm\"{u}ller-Tye potential model~\cite{Eichten:1995ch},
which corresponds to $\langle {\cal O}\rangle_{J/\psi}=0.387\:{\rm GeV^3}$, evaluated at the default factorization scale $\mu_\Lambda = 1$ GeV.

At $\sqrt{s}=10.58$ GeV, for $m_c=1.5$ GeV, we then obtain
$\sigma_\mathrm{int} = -0.761\:\text{fb}$,
$\sigma_\mathrm{nfr} = 0.0813\:\text{fb}$, respectively.
The interference term makes a sizable destructive contribution,  and the non-fragmentation part yields a negligible contribution.

\section{Higher-order radiative corrections in improved NRQCD factorization}
\label{Higher:order:prediction:improved:NRQCD}

\begin{figure}[h!]
    \centering
    \includegraphics[scale=1.0]{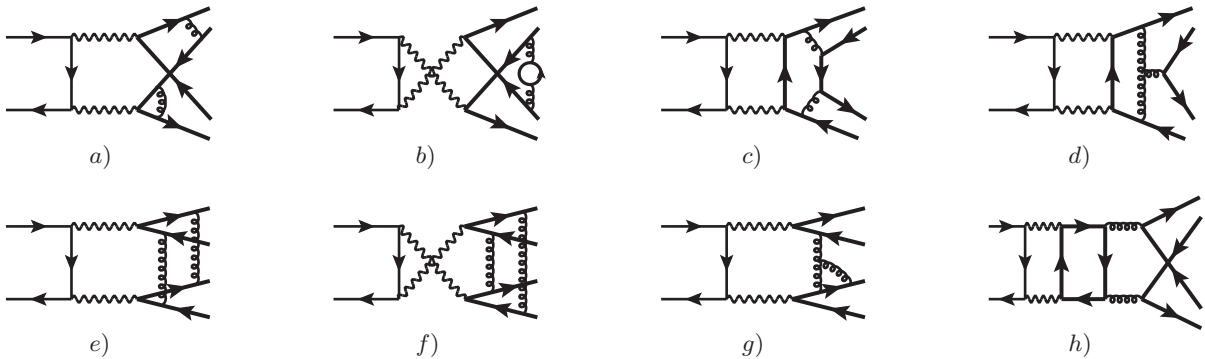}
    \caption{Some representative two-loop diagrams of non-fragmentation origin for $e^+e^-\to J/\psi+J/\psi$.}
    \label{Fig:nonfrag:diagrams:NNLO}
\end{figure}

We proceed to compute the ${\cal O}(\alpha_s)$ and ${\cal O}(\alpha_s^2)$  corrections to the unpolarized cross section.
In accordance with \eqref{Optimized:NRQCD:factorization:formula}, we only need consider those loop diagrams of non-fragmentation topology to avoid double counting.
We begin with the quark-level amplitude for
$e^+e^-\to \gamma^*\gamma^*\to c\bar{c}({}^3S_1^{(1)})+c\bar{c}({}^3S_1^{(1)})$.
At lowest order in $v$, we neglect the relative momentum in each $c\bar{c}$ pair prior to carrying out the
loop integration, which amounts to directly extracting the NRQCD SDCs from the hard loop region~\cite{Beneke:1997zp}.
We employ the dimensional regularization to regularize both UV and IR divergences.
About $24$ one-loop and $506$ non-vanishing two-loop diagrams of non-fragmentation type,
together with the corresponding amplitudes are generated by {\tt QGraf}/{\tt FeynArts}~\cite{Nogueira:1991ex,Hahn:2000kx}.
Some representative non-fragmentation types of one-loop and two-loop diagrams are shown in Fig.~\ref{Fig:nonfrag:diagrams:LO:NLO} and
Fig.~\ref{Fig:nonfrag:diagrams:NNLO}~\footnote{For simplicity, we have neglected those ``light-by-light''-type diagrams exemplified by
Fig.~\ref{Fig:nonfrag:diagrams:NNLO}$h)$, which generally yield very tiny contribution to
higher-order perturbative corrections in various quarkonium production and decay processes~\cite{Feng:2015uha,Sang:2015uxg,Feng:2017hlu,Sang:2020fql,Yang:2020pyh,Feng:2022vvk}.}.

We employ the covariant projector technique to ensure each $c\bar{c}$ pair to carry the intended $^3S_1^{(1)}$ quantum number.
We use the packages {\tt FeynCalc}/{\tt FormLink}~\cite{Mertig:1990an,Feng:2012tk} to
conduct the trace over Dirac and $SU(N_c)$ color matrices.
After the integration-by-parts (IBP) reduction with the aid of {\tt Apart}~\cite{Feng:2012iq} and {\tt FIRE}~\cite{Smirnov:2014hma},
we end up with about $2400$ two-loop master integrals (MIs).
We utilize the powerful package {\tt AMFlow}~\cite{Liu:2017jxz, Liu:2021wks, Liu:2022mfb, Liu:2022chg} to
compute the MIs with high numerical accuracy.
Note that the encountered two-loop diagrams have six eternal legs, which bear the genuine $2\to 4$ topology and
represent the cutting-edge problem in the area of multi-loop calculation.
The IBP reduction and computation of the MIs turn out to be rather time-consuming.

Performing the field-strength and mass renormalization, with two-loop expressions of $Z_2$ and $Z_m$ taken from
\cite{Broadhurst:1991fy},
and renormalizing the strong coupling constant under the $\overline{\rm MS}$ scheme to one-loop order,
we eliminate the UV divergences in the two-loop SDCs.
Nevertheless, the renormalized two-loop corrections to $\mathcal{C}_{\rm int}$ and $\mathcal{C}_{\rm nfr}$
still contain uncancelled single IR poles equal to ${\cal C}_{\rm int}^{(0)} \gamma_{J/\psi}$ and
to $2 {\cal C}_{\rm nfr}^{(0)} \gamma_{J/\psi}$, respectively.
This pattern is exactly what is required by NRQCD factorization for double $J/\psi$ production at ${\cal O}(\alpha_s)$,
as reflected in \eqref{C:int:nfr:param}.
The validity of NRQCD factorization in this process turns out to be highly nontrivial, as will be elaborated in the following section.
These IR poles can be factored into the NRQCD matrix element $\langle {\cal O}\rangle_{J/\psi}$
under the $\overline{\rm MS}$ prescription, which then becomes scale-dependent quantity. Note that the
$\gamma_{J/\psi}\ln \mu_{\Lambda}^2$ terms in \eqref{C:int:nfr:param} exactly cancel the $\mu_\Lambda$ dependence of the NRQCD matrix element,
so that the predicted cross section is independent of $\mu_{\Lambda}$.
Finally we are able to identify the desired non-logarithmic piece in the two-loop SDCs,
$\hat{c}_{\rm int}^{(2)}$ and $\hat{c}_{\rm nfr}^{(2)}$.

\begin{figure}[h!]
\centering
\includegraphics[scale=1.0]{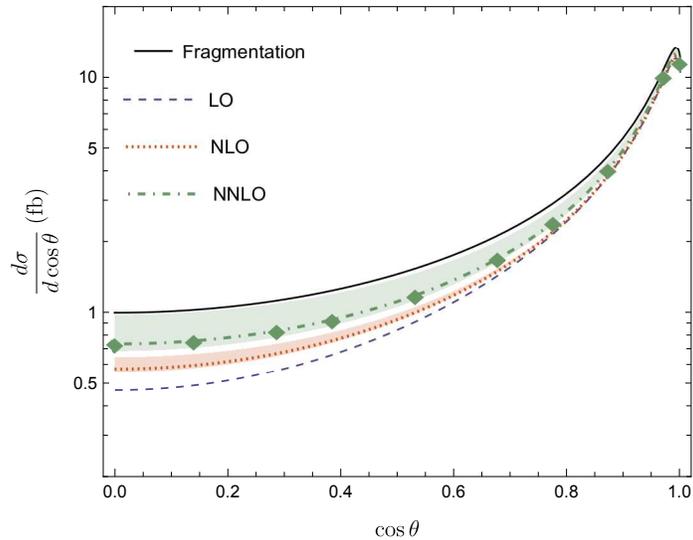}
\caption{Differential cross sections for $e^+ e^- \to J/\psi J/\psi$ against $\cos \theta$ at various perturbative accuracy
from improved NRQCD approach. We have fixed $\mu_{\Lambda} = 1\:\text{GeV}$, and taken the central value of $\mu_{R}$ to be $\sqrt{s}/2$.
The error bands of the NLO and NNLO predictions are estimated by sliding $\mu_R$ from $m_c$ to $\sqrt{s}$. }
\label{Plot:anuglar:distribution:double:Jpsi:Optimized:NRQCD}
\end{figure}

In our numerical analysis, we have fixed  $\sqrt{s}=10.58$ GeV, $M_{J/\psi} = 3.0969\:\text{ GeV}$, $f_{J/\psi} = 403\:\text{MeV}$,
$m_c=1.5$ GeV, and $\langle {\cal O}\rangle_{J/\psi}(\mu_\Lambda=1\:{\rm GeV})=0.387\:{\rm GeV^3}$.
The default value of $\mu_R$ is chosen as $\sqrt{s}/2$. We have varied $\mu_R$ from $m_c$ to $\sqrt{s}$ to estimate the theoretical uncertainty in
computing NLO and NNLO perturbative  corrections.
The package \texttt{RunDec}~\cite{Chetyrkin:2000yt} is utilized to computed the running QCD coupling to two-loop accuracy.
We have considered the two-loop corrections with ten sample points of $\cos \theta$  ranging from 0 to 1.

In Fig.~\ref{Plot:anuglar:distribution:double:Jpsi:Optimized:NRQCD} we plot the angular distribution of $J/\psi$
at various perturbative order from the improved NRQCD factorization. Due to the destructive interference between
the tree-level non-fragmentation amplitude (Fig.~\ref{Fig:nonfrag:diagrams:LO:NLO}$a)$ and $b)$) and fragmentaion amplitude,
the LO prediction is considerably smaller than the fragmentation cross section.
Nevertheless, both ${\cal O}(\alpha_s)$ and ${\cal O}(\alpha^2_s)$ corrections become positive in the improved NRQCD approach, which exhibit decent convergence behavior.
The modest positive NLO correction in our case is in sharp contrast with the substantial negative NLO correction
obtained from tradiational NRQCD approach~\cite{Gong:2008ce}.
In spite of large uncertainty, the finest prediction at NNLO accuracy
already gets quite close to the fragmentation prediction in \eqref{diff:X:section:photon:fragmentation}.
From Fig.~\ref{Plot:anuglar:distribution:double:Jpsi:Optimized:NRQCD}, one also sees that when the outgoing $J/\psi$ is collinear to the
electron beam direction, the fragmentation contribution dominates the cross section. As $\theta$ deviates from 0, the interference term
starts to play some notable role. The non-fragmentation term appears to be insignificant in the entire range of $\theta$.

\begin{figure}[H]
\centering
\includegraphics[width=0.31\textwidth]{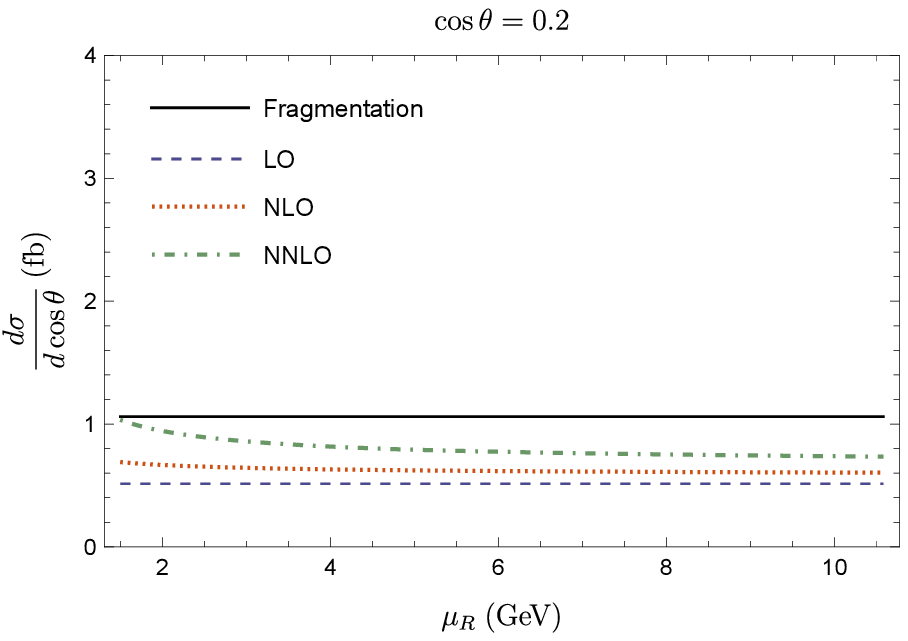} \quad
\includegraphics[width=0.31\textwidth]{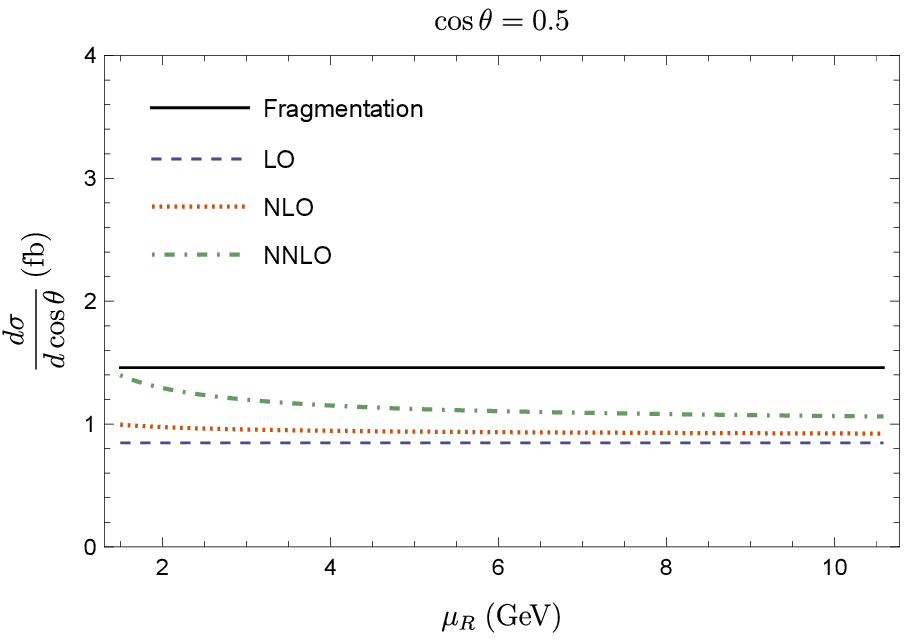} \quad
\includegraphics[width=0.31\textwidth]{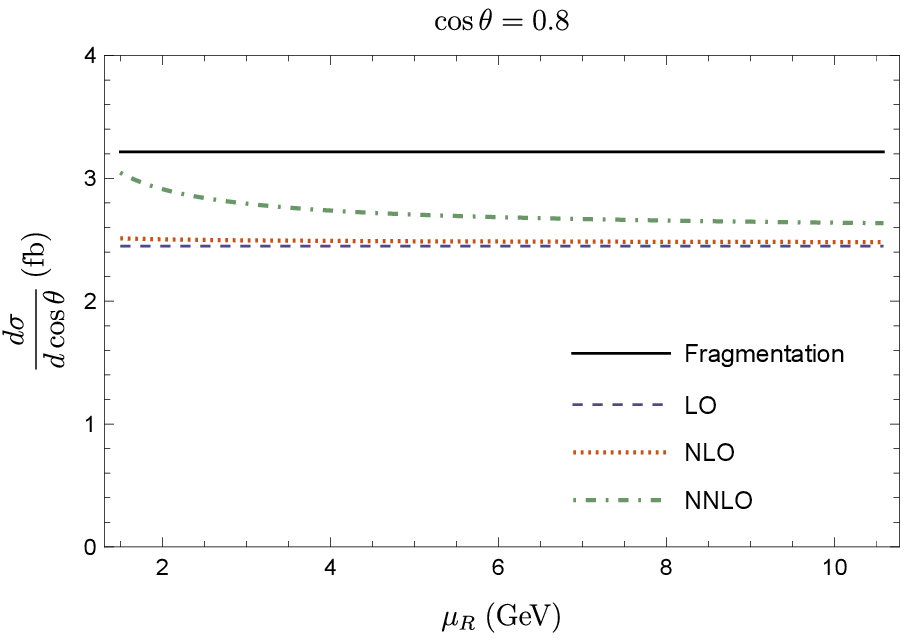}
\caption{Renormalization scale dependence of the differential cross sections at various perturbative accuracy,
with three typical values of $\theta$. }
\label{muR:dependence:Diff:X:Section:Optimized}
\end{figure}

In Fig.~\ref{muR:dependence:Diff:X:Section:Optimized} we also show the the $\mu_R$ dependence of the differential cross section
at various perturbative accuracy, for $\cos\theta=0.2,\, 0.5, \, 0.8$, respectively.
Interestingly, we observe that the NLO predictions exhibit a very flat $\mu_R$ dependence in the improved NRQCD approach, while the NNLO
predictions possess a reasonably flat $\mu_R$ dependence except in the lower $\mu_R $ end.
It is mainly due to the good convergence behavior of the perturbative expansion.
In particular, as indicated in \eqref{C:int:nfr:param}, the coefficients of $\beta_0\ln \mu_R^2$ are proportional to the interference contribution
${\cal C}_{\text{int}}$ and non-fragmentation contribution ${\cal C}_{\text{nfr}}$, which are considerably smaller than ${\cal C}_{\rm fr}$.

\begin{table}[!htbp]\small
        \centering
    \setlength{\tabcolsep}{6pt}
    \renewcommand{\arraystretch}{1.6}
    {
        \begin{tabular}{|c|c|c|c|c|c|c|c|}
            \hline
 $\cos\theta$   & \makecell[c]{$\mathcal{C}_{\text{fr}}$\\$(\text{GeV}^{-4})$} & \makecell[c]{${\cal C}_{\rm int}^{(0)}$\\$(\text{GeV}^{-4})$} &
 $\hat{c}_{\rm int}^{(1)}$ & $\hat{c}_{\rm int}^{(2)}$ & \makecell[c]{${\cal C}_{\rm nfr}^{(0)}$\\$(\text{GeV}^{-4})$} & $\hat{c}_{\rm nfr}^{(1)}$ & $\hat{c}_{\rm nfr}^{(2)}$
\\\hline
 $ 0.999 $ &
$4.163$ & $ -0.334 $ & $ -3.62 $ & $ -71.75 $ & $ 0.006 $ & $ -7.42 $ & $-143.174+ 42.974=-100.20$
\\\hline
 $ 0.970 $ &
$3.646$ & $ -0.242 $ & $ -1.34 $ & $ -76.57 $ & $ 0.007 $ & $ -6.33 $ & $-146.117+ 37.424=-108.69$
\\\hline
 $ 0.872 $ &
$1.573$ & $ -0.193 $ & $ -0.73 $ & $ -80.64 $ & $ 0.008 $ & $ -5.07 $ & $-152.144+ 25.321=-126.82$
\\\hline
 $ 0.775 $ &
$0.988$ & $ -0.176 $ & $ -1.27 $ & $ -81.77 $ & $ 0.010 $ & $ -5.11 $ & $-155.633+ 19.124=-136.51$
\\\hline
 $ 0.677 $ &
$0.722$ & $ -0.164 $ & $ -1.85 $ & $ -82.00 $ & $ 0.011 $ & $ -5.49 $ & $-157.716+ 15.969=-141.75$
\\\hline
 $ 0.531 $ &
$0.522$ & $ -0.152 $ & $ -2.58 $ & $ -81.67 $ & $ 0.012 $ & $ -6.15 $ & $-159.349+  14.092 =-145.26$
\\\hline
$ 0.384 $ &
$0.422$ & $ -0.143 $ & $ -3.12 $ & $ -81.08 $ & $ 0.012 $ & $ -6.73 $ & $-160.032+  13.777=-146.26$
\\\hline
 $ 0.287 $ &
$0.383$ & $ -0.139 $ & $ -3.38 $ & $ -80.71 $ & $ 0.012 $ & $ -7.03 $ & $-160.222+  13.898=-146.32$
\\\hline
 $ 0.140 $ &
$0.350$ & $ -0.135 $ & $ -3.63 $ & $ -80.31 $ & $ 0.012 $ & $ -7.32 $ & $-160.324+ 14.160 =-146.16$
\\\hline
$ 0 $ &
$0.340$ & $ -0.133 $ & $ -3.70 $ & $ -80.17 $ & $ 0.012 $ & $ -7.41 $ & $-160.341+ 14.271=-146.07$
\\\hline
\end{tabular}
}
\label{Table:SDCs:optimized:NRQCD}
\caption{Numerical values of various SDCs in \eqref{C:int:nfr:param} through ${\cal O}(\alpha^2_s)$ for ten different values of $\cos\theta$. }
\end{table}

For reader's convenience, we have tabulated in Table~\ref{Table:SDCs:optimized:NRQCD} all the SDCs through ${\cal O}(\alpha^2_s)$ that
enter \eqref{C:int:nfr:param}, for ten different values of $\cos\theta$.
In the last column in Table~\ref{Table:SDCs:optimized:NRQCD}, we split $\hat{c}_{\rm nfr}^{(2)}$ into two terms which have different
sources, the inner product between the two-loop non-fragmentation amplitude and the tree-level non-fragmentation amplitude,
and the absolute square of the one-loop non-fragmentation amplitude, respectively. We clearly see that only ${\cal C}_{\rm fr}$ exhibits
strong angular dependence. The non-fragmentation SDCs ${\cal C}_{\rm nfr}$ are always insignificant for entire range of $\theta$.
The importance of the interference term ${\cal C}_{\rm int}$ increases as $\cos\theta$ decreases. Since the SDCs $\hat{c}_{\rm int}^{(1,2)}$ have negative sign,
including higher-order radiative corrections turn to dilute the destructive interference effect.

\begin{table}
\setlength{\tabcolsep}{6pt}
\renewcommand{\arraystretch}{1.6}
\begin{tabular}{ccccc}
\hline
  $\sigma$ (fb) & Fragmentation & LO & NLO & NNLO
\\\hline\hline
  Optimized NRQCD &
  \multirow{2}{*}{$2.52$} & $1.85$ & $1.93^{+0.05}_{-0.01}$ & $2.13^{+0.30}_{-0.06}$
\\
  Traditional NRQCD &
  & $6.12$ & $1.56^{+0.73}_{-2.95}$ & $-2.38^{+1.27}_{-5.35}$ \\
\hline
\end{tabular}
\caption{Integrated cross section of $e^+e^-\to J/\psi J/\psi$ at various perturbative accuracy.
The uncertainties are estimated by varying $\mu_R$ from $m_c$ to $\sqrt{s}$. }
\label{Integrated:Cross:Sections:B:factory}
\end{table}

Finally in Table~\ref{Integrated:Cross:Sections:B:factory} we enumerate our predictions at various perturbative accuracy
for the integrated cross section at $\sqrt{s}=10.58$ GeV.
The NNLO prediction from the improved NRQCD approach, $2.13^{+0.30}_{-0.06}$ fb,
is already close to the fragmentation contribution, 2.52 fb.
In contrast with the negative cross section predicted by the standard NRQCD approach,
we believe that our prediction based on optimized NRQCD approach
is robust and reliable~\footnote{We caution that, since we have not included the uncertainty inherent in charm quark mass
and NRQCD matrix element, the true error bar should be somewhat
greater than what is listed in Table~\ref{Integrated:Cross:Sections:B:factory}.
$m_c$ is often assumed to vary from 1.4 to 1.5 GeV. The value of the NRQCD matrix element often varies in different work.
Concretely speaking, the value of $\langle {\cal O}\rangle_{J/\psi}$ is taken to be
$0.335\pm0.024\;{\rm GeV^3}$~\cite{Bodwin:2002fk,Bodwin:2002kk},  $0.482\pm0.049\:{\rm GeV^3}$~\cite{Bodwin:2006yd}, $0.457\:{\rm GeV^3}$~\cite{Gong:2008ce},
$0.440^{+0.067}_{-0.055}\:{\rm GeV^3}$ or $0.436^{+0.065}_{-0.054}\:{\rm GeV^3}$ (depending on the value of $m_c$)~\cite{Fan:2012dy}.
Nevertheless, since the bulk of the cross section arises from the fragmentation contribution, we anticipate that the uncertainty
caused by $m_c$ and $\langle {\cal O}\rangle_{J/\psi}$ in the improved NRQCD approach
is much less than that from the traditional NRQCD approach.}.

To date the \texttt{Belle } and the \texttt{Belle 2} experiments have accumulated about 1500 ${\rm fb}^{-1}$ data, so we expect about $3105\sim3645$ exclusive
double $J/\psi$ events. Taking into account ${\cal B}(J/\psi\to l^+l^-) = 12\%$, about $45\sim52$ four-lepton events from double $J/\psi$ can be produced.
Assuming 40\% reconstruction efficiency, we expect about $18\sim21$ signal events may be reconstructed.
Concerning the potentially copious background events, it might be challenging to unambiguously establish
the double $J/\psi$ signals based on the current dataset.
Nevertheless, with the designed 50 ${\rm ab}^{-1}$ integrated luminosity at \texttt{Belle 2},
it seems that the observation prospects of exclusive double $J/\psi$ production is  promising in the foreseeable future~\footnote{It is interesting to note that, our NNLO prediction for $\sigma(e^+e^-\to \gamma^*\gamma^*\to J/\psi J/\psi)$
is actually greater than the NNLO predictions for $\sigma(e^+e^-\to \gamma^*\to J/\psi+\chi_{c1})= {0.87}^{+0.19}_{-0.29}$ fb and
$\sigma(e^+e^-\to \gamma^*\to J/\psi+\chi_{c2})=  {0.73}^{+0.17}_{-0.27}$ fb at $B$ factories~\cite{Sang:2022kub}.}.

\section{Nontrivial examination of NRQCD factorization at two-loop}
\label{two:loop:NRQCD:examination}

The validity of NRQCD factorization proves to be rather intriguing at two-loop order.
We pick up a specific non-fragmentation type of two-loop diagram, Fig.~\ref{Fig:nonfrag:diagrams:NNLO}$e)$, to illustrate this point.
This Abelian two-loop diagram, together with additional 17 similar diagrams, can be obtained by dressing the photon-fragmentation diagram in
Fig.~\ref{diagrams:2Jpsi:from:photon:fragmentation}$a)$ with two gluon exchanges between two $c\bar{c}$ pairs.
Note Fig.~\ref{Fig:nonfrag:diagrams:NNLO}$g)$ which contains non-abelian three-gluon vertex
gives a vanishing contribution since each of the $c\bar{c}$ pairs must be color-singlet. Similarly, there are also 18 two-loop non-fragmentation
diagrams that can be obtained by dressing the crossed photon-fragmentation diagram in
Fig.~\ref{diagrams:2Jpsi:from:photon:fragmentation}$b)$ with all possible two gluon exchange.

When $\cos{\theta} \neq0$, Fig.~\ref{Fig:nonfrag:diagrams:NNLO}$e)$ exhibits severe IR divergences which start at
$1/\epsilon_{\rm IR}^3$, with complex-valued coefficients. This leading divergence might arise from the overlap
between the Sudakov double IR pole from the electron-photon loop and the soft pole arising
from another quark-gluon loop.
The subleading IR poles have more complicated origin.
Had these IR poles not been cancelled, the NRQCD factorization would break down, since these IR divergences cannot be
affiliated with an individual quarkonium, let alone to be factored into the respective NRQCD matrix element.
Fortunately, all the IR poles, from ${\cal O}(1/\epsilon_{\rm IR}^3)$ to ${\cal O}(1/\epsilon_{\rm IR})$,
exactly cancel upon summing 18 two-gluon exchange diagrams together.

A simplifying situation arises as $\cos{\theta}=0$, that the leading IR pole starts at order~$1/\epsilon_{\rm IR}^2$.
The origin of this double IR pole is clear, which stems from the loop regions where both gluons become simultaneously soft.
After making eikonal approximation, we find
\begin{align}
    {\cal M}^{{\rm Fig}.4e)} \bigg|_{\theta={\pi\over 2}} = {1\over \epsilon^2_{\rm IR}}
    \frac{C_F\,\alpha^2}{2 N_c} \frac{(m_c^2+2 \mathbf{P}^2)^2}{16 \mathbf{P}^2
    (4m_c^2+\mathbf{P}^2)}\left( \ln{\frac{1+\beta}{1-\beta}}-i \pi \right)^2
    {\cal M}_{{\rm fr},0}^{{\rm Fig.}1a)}\bigg|_{\theta={\pi\over 2}} + {\cal O}(1/\epsilon_{\rm IR}),
\end{align}
where $|\mathbf{P}|$ denotes the magnitude of the $J/ \psi$ momentum.
The same double IR pole is shared by other 5 diagrams, and the coefficients of the double IR pole in other 12 diagrams have an extra $-1/2$ factor.
Summing up 18 diagrams, these double IR poles exactly cancel.
The pattern of the cancelation of the single IR poles become much more involved.
To summarize, the cancellation of IR poles among these 18 two-gluon exchange diagrams implies that the
NRQCD factorization is fulfilled in a highly nontrivial manner.

\section{Summary}
\label{Summary}

The exclusive double $\rho^0$ production from $e^+e^-$ annihilation has been observed by \texttt{BaBar} experiment
about two decades ago. However, to date the very clean $e^+ e^- \to J/\psi J/\psi$ process has not been discovered yet.
This exclusive double vector charmonium production is theoretically interesting,
which has to proceed through two virtual photon annihilation due to the $C$-even feature of the final state.
Complementary to the well-studied double charmonium production processes $e^+e^-\to J/\psi+\eta_c (\chi_{cJ})$,
a better understanding of this double $J/\psi$ production process can enrich our knowledge about
charmonium production mechanism and test the applicability of NRQCD factorization approach.

The significant negative NLO perturbative correction may indicate that the production rate is too small for this double-$J/\psi$ production process
to be observed at $B$ factory. Needless to say, an accurate theoretical account for the cross section
offers crucial guidance for experimental search of this process.

In this work we calculate the ${\cal O}(\alpha^2_s)$ correction to this process for the first time.
It is found that the standard NRQCD approach would result in a substantially negative NNLO perturbative correction,
and inevitably lead to the unphysical negative cross section.
The symptom may be attributed to the observation that,
the bulk contribution of the fixed-order radiative corrections actually stems from the perturbative corrections
to the $J/\psi$ decay constant, which are also negative and significant, at least at first few orders
in the strong coupling constant.
Motivated by this observation, we implement an improved NRQCD factorization approach,
in which the amplitude is split into the photon-fragmentation piece and the non-fragmentation piece.
The fragmentation-induced production rate can be predicted unambiguously with the measured $J/\psi$ decay constant as input.
The interference part and the non-fragmentation part are then computed through NNLO in $\alpha_s$ and at lowest order in velocity.
In this optimized scheme, we find that both the ${\cal O}(\alpha_s)$ and ${\cal O}(\alpha^2_s)$ corrections in the interference part
become positive and exhibit a reasonable  convergence pattern.
The non-fragmentation part turns out to be insignificant numerically.
Our most accurate prediction in the optimized NRQCD approach is $\sigma(e^+ e^- \to J/\psi+J/\psi)=2.13^{+0.30}_{-0.06}$ fb  at $\sqrt{s}=10.58$ GeV,
which is quite close to the fragmentation prediction. We believe that this NNLO prediction is much more
meaningful and trustworthy than that from the traditional NRQCD approach.
It is worth making a more reliable error estimation by including the uncertainty affiliated with the charm quark mass and NRQCD matrix element
in the future work.

At present, it looks challenging, but not impossible, to unambiguously establish the double $J/\psi$ signals based on the current 1500 ${\rm fb}^{-1}$
data accumulated in \texttt{Belle} and \texttt{Belle 2}.
With the projected 50 ${\rm ab}^{-1}$ full data set at \texttt{Belle 2}, the observation prospect of the exclusive
double $J/\psi$ production process looks very bright.

\begin{acknowledgments}
We are grateful to Xiao Liu and Cheng-Ping Shen for useful discussions.
The work of W.-L. S. is supported by the NNSFC Grant No. 11975187.
The work of F.~F. is supported by the NNSFC Grant No. 12275353 and No. 11875318.
The work of Y.~J., Z.~M., J.~P. and J.-Y.~Z is supported in part by the NNSFC Grants No.~11925506, No.~12070131001 (CRC110 by DFG and NSFC).
\end{acknowledgments}

\appendix

\section{Perturbative corrections to $e^+e^-\to J/\psi+J/\psi$ in traditional NRQCD framework}
\label{appendix:A}

For the sake of completeness, in this appendix we also show the results for the perturbative corrections $e^+e^-\to J/\psi+J/\psi$ from the standard NRQCD factorization approach.
We take the shortcut to arrive at \eqref{Conventional:NRQCD:factorization:formula} directly from in improved NRQCD factorization predictions.
Setting $M_{J/\psi}=2 m_c$ in (\ref{Optimized:NRQCD:factorization:formula}), and expressing the decay constant $f_{J/\psi}$ in terms of $\langle {\cal O}\rangle_{J/\psi}$
in accordance with (\ref{decay:constant:NRQCD:fac}), we can rewrite the differential cross section as
\beq
\frac{d\sigma}{d\cos\theta}= {1\over 2s} {\beta \over 16\pi}\frac{e^8e_c^4}{4} {\cal F}^{(0)}
\bigg[
1+\frac{\alpha_s}{\pi} f^{(1)}+\bigg(\frac{\alpha_s}{\pi}\bigg)^2
\bigg(f^{(1)}\frac{\beta_0}{4}\ln\frac{\mu_R^2}{m_c^2}+4\gamma_{J/\psi}\ln\frac{\mu_\Lambda^2}{m_c^2}
+f^{(2)}\bigg)
\bigg]\frac{\langle {\cal O}\rangle_{J/\psi}^2}{m_c^2},
\label{F:traditional:NRQCD:param}
\eeq
where
\begin{subequations}
\bqa
{\cal F}^{(0)}&=&\mathcal{C}_{\text{fr}}+ {\cal C}_{\rm int}^{(0)}+ {\cal C}_{\rm nfr}^{(0)},
\\
f^{(1)}&=& \tilde{f}^{(1)}+\frac{2\mathfrak{f}^{(1)}}{{\cal F}^{(0)}}{\cal C}_{\text{int}}^{(0)}+ \frac{{\cal C}_{\text{int}}^{(0)}
\hat{c}_{\rm int}^{(1)}+{\cal C}_{\text{nfr}}^{(0)} \hat{c}_{\rm nfr}^{(1)}}{{\cal F}^{(0)}},
\label{f1:reexpand}
\\
f^{(2)}&=& \tilde{f}^{(2)}+\frac{(\mathfrak{f}^{(1)})^2}{{\cal F}^{(0)}}{\cal C}_{\text{int}}^{(0)}+\frac{2\mathfrak{f}^{(2)}}{{\cal F}^{(0)}}{\cal C}_{\text{int}}^{(0)} + \frac{2\mathfrak{f}^{(1)}}{{\cal F}^{(0)}}{\cal C}_{\text{int}}^{(0)}\hat{c}_{\rm int}^{(1)}+\frac{{\cal C}_{\text{int}}^{(0)} \hat{c}_{\rm int}^{(2)}+{\cal C}_{\text{nfr}}^{(0)} \hat{c}_{\rm nfr}^{(2)}}{{\cal F}^{(0)}}.
\label{f2:reexpand}
\eqa
\end{subequations}
Here we have singled out the fragmentation contributions in each SDC,
which are encapsulated in the coefficients $\tilde{f}^{(1)}$ and $\tilde{f}^{(2)}$:
\beq
\tilde{f}^{(1)}=\frac{4\mathfrak{f}^{(1)}}{ {\cal F}^{(0)} } \mathcal{C}_{\text{fr}},  \qquad
\tilde{f}^{(2)}=\frac{6\left(\mathfrak{f}^{(1)}\right)^2+4\mathfrak{f}^{(2)} }{{\cal F}^{(0)}} \mathcal{C}_{\text{fr}},
\eeq
defined by the fragmentation contribution multiplied by the radiative corrections to the $J/\psi$ decay constant.

\begin{figure}[H]
\centering
\includegraphics[scale=1.0]{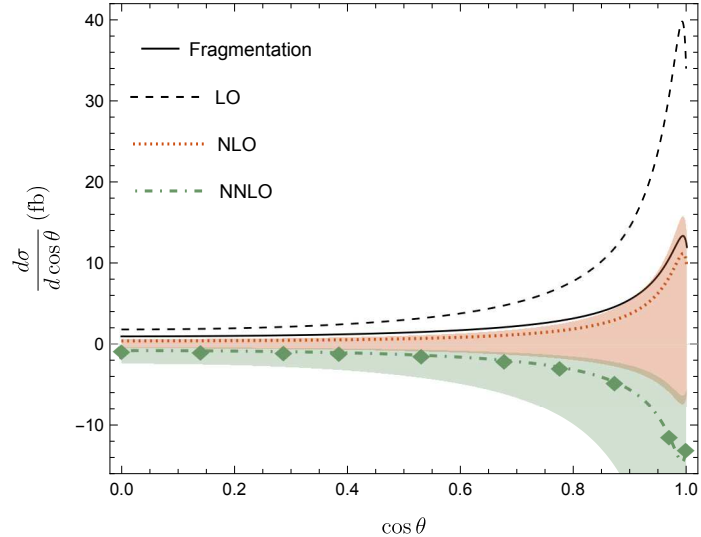}
\caption{Differential cross sections of $e^+ e^- \to J/\psi J/\psi$ against $\cos \theta$ at various perturbative accuracy
from {\it conventional} NRQCD approach. We have fixed $\mu_{\Lambda} = 1\:\text{GeV}$, and taken the central value of $\mu_{R}$ to be $\sqrt{s}/2$.
The error bands of the NLO and NNLO predictions are estimated by sliding $\mu_R$ from $m_c$ to $\sqrt{s}$. }
\label{Plot:anuglar:distribution:double:Jpsi:Traditional:NRQCD}
\end{figure}

In Fig.~\ref{Plot:anuglar:distribution:double:Jpsi:Traditional:NRQCD} we plot the angular distribution of $J/\psi$
at various perturbative order within the traditional NRQCD factorization framework.
The LO prediction is considerably greater than the fragmentation contribution.
Nevertheless, both ${\cal O}(\alpha_s)$ and ${\cal O}(\alpha^2_s)$ corrections
significantly decrease the LO prediction. We have confirmed the significant negative ${\cal O}(\alpha_s)$
correction first discovered by Gong and Wang~\cite{Gong:2008ce}.
Remarkably, the NNLO perturbative correction is also very substantial, which brings the predicted cross section down
to unphysical, negative value.  This pathetic symptom can also be clearly seen in the integrated cross sections in
Table~\ref{Integrated:Cross:Sections:B:factory}.

\begin{figure}[H]
\centering
\includegraphics[width=0.31\textwidth]{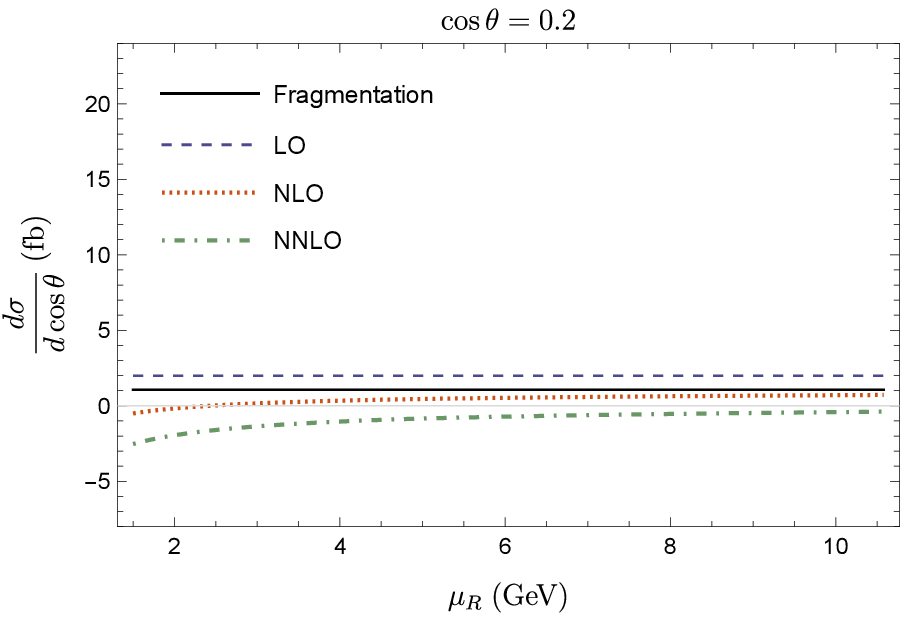} \quad
\includegraphics[width=0.31\textwidth]{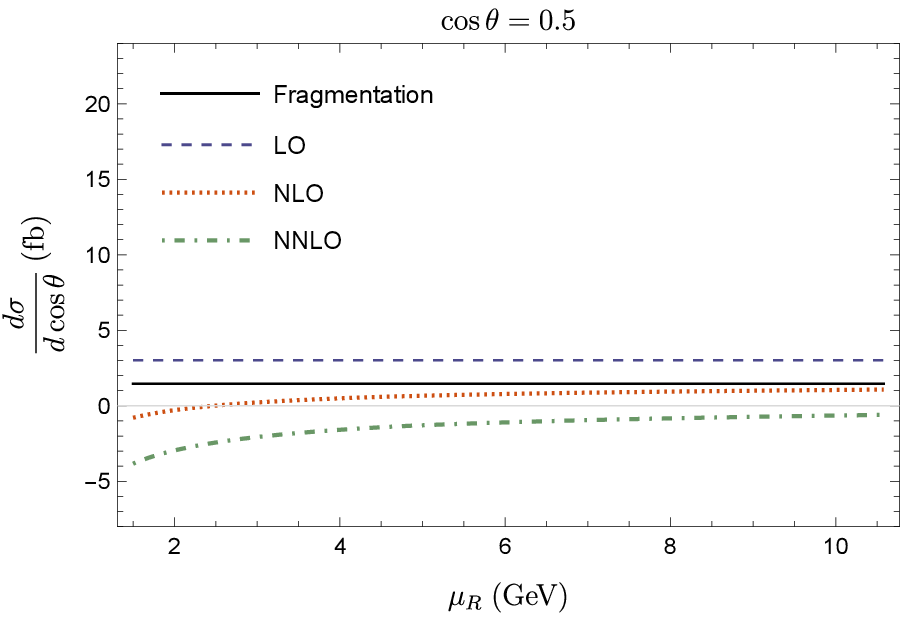} \quad
\includegraphics[width=0.31\textwidth]{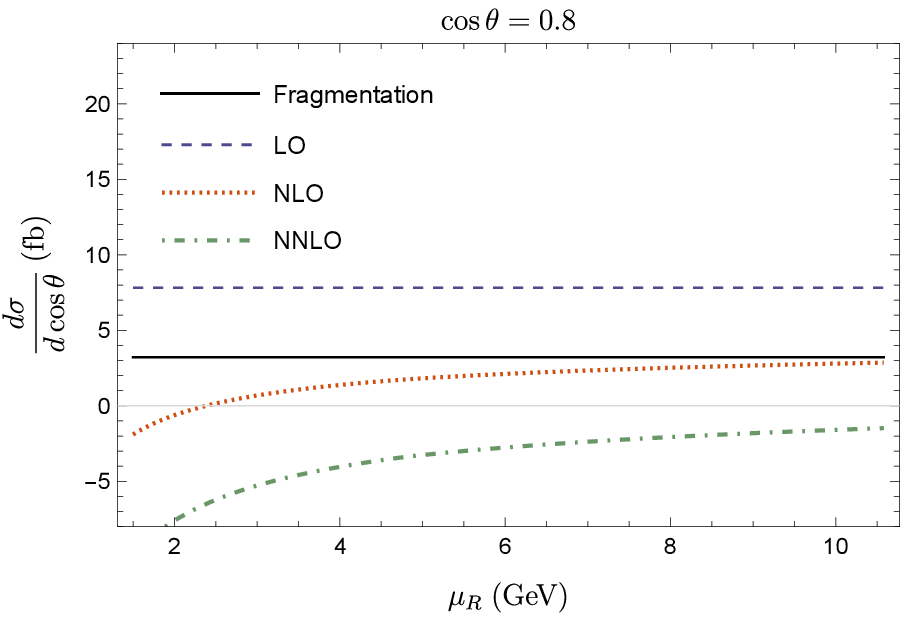}
\caption{$\mu_R$-dependence of the differential cross sections at various perturbative accuracy,
with three typical values of $\theta$. }
\label{muR:dependence:Diff:X:Section:Traditional}
\end{figure}

In Fig.~\ref{muR:dependence:Diff:X:Section:Traditional} we also show that renormalization scale dependence of
the differential cross sections at various perturbative accuracy, for some typical values of $\cos\theta$.
In contrast to the mild $\mu_R$ dependence in the improved NRQCD prediction as shown in Fig.~\ref{muR:dependence:Diff:X:Section:Optimized},
the NLO and NNLO predictions in the traditional NRQCD approach exhibit stronger $\mu_R$ dependence.

\setcounter{equation}{0}
\renewcommand\theequation{A\arabic{equation}} 

\begin{table}[!htbp]\small
        \centering
    \setlength{\tabcolsep}{6pt}
    \renewcommand{\arraystretch}{1.6}
    {
        \begin{tabular}{|c|c|c|c|c|c|}
            \hline
 $\cos\theta$   &  ${\cal F}^{(0)}$ & $f^{(1)}$ & $\tilde{f}^{(1)}$& $f^{(2)}$ &  $\tilde{f}^{(2)}$   \\
 \hline
0.999 & 4.76 & -10.78 & -11.40 & -130.394-0.123=-130.52 & -139.65
\\
\hline
 0.970 & 4.14 & -10.89 & -11.27 & -129.370-0.172=-129.54 & -138.08
\\
\hline
 0.872 & 1.60 & -11.19 & -11.90 & -126.868-0.670=-127.54 & -145.70
\\
 \hline
 0.775 &  0.94 &  -11.37  &  -12.55  & -124.802-1.429=-126.23 & -153.77
\\
\hline
0.677 &  0.65 &  -11.47  &  -13.20  & -123.136-2.354=-125.49 & -161.67
\\
 \hline
 0.531 &  0.43 &  -11.51 &  -14.11  & -121.319-3.853=-125.17 & -172.76
 \\
\hline
0.384 & 0.33  &  -11.48 & -14.88 & -120.211-5.246=-125.46 & -182.30
\\
\hline
 0.287 &  0.29 &  -11.45  &  -15.30  & -119.781-6.006=-125.79 & -187.39
 \\
\hline
 0.140 &  0.26 &  -11.40 &  -15.72 & -119.465-6.779=-126.24 & -192.57
\\
 \hline
 0 &  0.25 &  -11.38 &  -15.86 & -119.364-7.032=-126.40 & -194.27
 \\
 \hline
\end{tabular}
    }
\label{Table:SDCs:Traditional:NRQCD}
\caption{Numerical values of various SDCs in \eqref{F:traditional:NRQCD:param} through ${\cal O}(\alpha^2_s)$ for ten different values of $\cos\theta$.}
\end{table}

For reader's convenience, in Table~III we summarize various SDCs through ${\cal O}(\alpha^2_s)$
in the traditional NRQCD framework.  In the last column in Table~III, we split $f^{(2)}$ into two pieces
which have different origin: the first piece includes the fragmentation and interference contributions, and the second term represents
the non-fragmentation contribution ${\cal C}_{\rm nfr}^{(0)} \hat{c}_{\rm nfr}^{(2)}/{\cal F}^{(0)}$ in \eqref{f2:reexpand}.
Note only the LO SDC ${\cal F}^{(0)}$ exhibits strong angular dependence,
while the normalized higher-order SDCs $f^{(1,2)}$ have mild angular dependence.

The striking observation is that in a wide range of $\cos\theta$, the order-$\alpha_s$ SDC $f^{(1)}$ is well saturated by
$\tilde{f}^{(1)}$, and $f^{(2)}$ is reasonably approximated by $\tilde{f}^{(2)}$.  Since $\tilde{f}^{(i)}$ ($i=1,2$)
characterize the perturbative corrections to the $J/\psi$ decay constant,
the bulk contribution of the fixed-order perturbative corrections to the double-$J/\psi$ cross section
in standard NRQCD factorization actually arises from the perturbative corrections to the $J/\psi$ decay constant.
Because the perturbative corrections to $f_{J/\psi}$ in \eqref{decay:constant:NRQCD:fac}
become increasingly negative, it is not surprising to encounter the negative NLO and NNLO radiative corrections
to $\sigma(e^+e^-\to J/\psi J/\psi)$ in standard NRQCD factorization.
In contrast, in our improved NRQCD approach, the predicted cross section at each perturbative order is always positive.

We can also readily understand why the higher-order contributions in the traditional NRQCD approach exhibit stronger $\mu_R$
dependence than that in the improved NRQCD approach. The key reason is that the perturbative convergence in the former is much worse
than the latter.


\end{document}